
\font\elvrm=cmr10  scaled\magstep1
\font\elvmi=cmmi10  scaled\magstep1
\font\elvsy=cmsy10  scaled\magstep1
\font\elvex=cmex10  scaled\magstep1
\font\elvit=cmti10  scaled\magstep1
\font\elvsl=cmsl10  scaled\magstep1
\font\elvbf=cmbx10  scaled\magstep1

\font\ninerm=cmr9         \font\eightrm=cmr8          \font\sixrm=cmr6
         \font\eighti=cmmi8          \font\sixi=cmmi6
        \font\eightsy=cmsy8         \font\sixsy=cmsy6

\newcount\textno
\newcount\scriptno
\global\textno=10
\global\scriptno=7


\textno=12\scriptno=7
\normalbaselineskip=13dd  
\textfont0=\elvrm  \scriptfont0=\eightrm  \scriptscriptfont0=\sixrm
\textfont1=\elvmi  \scriptfont1=\eighti  \scriptscriptfont1=\sixi
\textfont2=\elvsy  \scriptfont2=\eightsy  \scriptscriptfont2=\sixsy
\textfont3=\elvex  \scriptfont3=\elvex    \scriptscriptfont3=\elvex
\textfont4=\elvit  \textfont5=\elvsl      \textfont6=\elvbf
\def\rm{\fam0\elvrm}                      \def\it{\fam4\elvit}
\def\sl{\fam5\elvsl}                      \def\bf{\fam6\elvbf}
\setbox\strutbox=\hbox{\vrule height8.5dd depth4dd width0dd}
\normalbaselines\rm
{\vskip-2cm
\headline={\tt 
$\backslash$ 
$\backslash$ 
}\vskip2cm}

\vsize19.6cm \hsize12.8cm \baselineskip=7mm
\voffset2.5pc\hoffset2.5pc

\nopagenumbers
\def\rightheadline{{\eightrm\hfil
\hfil}\folio}
\def\leftheadline{\folio\eightrm\hfil
\hfil}
\headline={\vbox to 0pt{\vskip-1.82pc\line{\vbox to 8.5pt{}
\ifodd\pageno\rightheadline \else\leftheadline\fi}\vss}}

\def\ipsi{{\psi\atop\qquad}\hskip-23pt\to}

\centerline{\bf Efficacy of non-locality theorems ``without inequalities''}
\centerline{\bf for pairs of spin-$1\over 2$ particles}
\vskip2pc
\centerline{\sl Giuseppe Nistic\`o}
\par\centerline{\it Dipartimento di Matematica, Universit\`a della Calabria}
\par\centerline{\it 87036 -- Arcavacata RENDE (CS) Italy}
\par\centerline{\it tel. (++)39(+)984 493259, fax (++)39(+)984 401186}
\par\centerline{e-mail gnistico@unical.it}
\vskip2pc
\noindent
{\ninerm
We argue that for a system of two spin-$1\over 2$ particles the recent
theorems without inequalities, which show non-locality
of quantum theory,
fail in proving
non-locality of any empirically valid
theory sharing a set of correlations with
quantum theory.
In this case, a Greenberger-Horne-Shimony-Zeilinger
argument cannot work.}
\par\noindent
{\it keywords}: nonlocality, quantum theory.
\vskip1pc\noindent
The subject of the present paper are the insights provided by
the non-locality theorems without inequalities for two spin-$1\over 2$
particles. The theorem proposed by Hardy [1][2], in particular,
has gained much interest in the literature on this subject [3]-[5].
Contrary to Bell's theorem [6],
it does not make use of inequalities and
works for almost all entangled states.
In this letter we show that such new theorems, once proved that quantum
theory is not a local and realistic theory, cannot extend this
negative result to any theory which share only correlations, and not statistics
with quantum theory. This stronger non-locality proof
is attained by Greenberger,
Horne, Shimony and Zeilinger ({\sl GHSZ}, from now on) for a system consisting of at least three
particles [7]. We show that
the method of GHSZ cannot apply in the case of pairs
of two-level particles.
\vskip1pc
Now we present an equivalent reformulation of Hardy's argument.
It involves two spin-$1\over 2$, space-like separated
particles: particle 1  and particle 2.
By $S_k({\bf n})$ ($k=1,2$) we denote the 1-0 observable which assumes
value 1 (resp., 0) when the spin of particle $k$ in direction
$\bf n$ is ${1\over 2}\hbar$ (resp., $-{1\over 2}\hbar$). Four
particular directions
${\bf n}_1$, ${\bf n}_2$, ${\bf n}_3$ and
${\bf n}_4$ are also considered, such that
${\bf n}_j$ and ${\bf n}_{j+2}$ are not parallel.
The ingredients of Hardy's theorem are the following definition
of element of reality and statements $(i)$, $(ii)$ and $(iii)$.
\vskip1pc\noindent
D{\ninerm EFINITION.} 1. --
\sl
We say that an observable $S$ is an element of reality [equal to $s$]
if a value [$s$] is assigned to $S$, albeit unknown, such that a measurement
of $S$ would yield that particular value [$s$].
\rm
\vskip1pc\noindent
\item{$i$)} {\sl Principle of locality and reality}
\item{}
Let $S_1$ and $S_2$ be two physical magnitudes which
are measurable in two space-like separated regions.
If a measurement of $S_1$ allows the prediction of the outcome of a measurement of
$S_2$, then $S_2$ is an element of reality, {\sl no matter whether
$S_1$ is actually measured or not.}
\vskip1pc\noindent
\item{$ii$)}{\sl Three correlations}
\item{}
1)\quad
$S_1({\bf n}_1)\;\rightarrow\; S_2({\bf n}_2)$ ;
by this formula it is meant that
if $S_1({\bf n}_1)$ and $S_2({\bf n}_2)$ are measured,
then the outcome 1 for $S_1({\bf n}_1)$ implies that also the outcome
of $S_2({\bf n}_2)$ is 1.
\item{}
2)\quad
$S_2({\bf n}_2)\rightarrow S_1({\bf n}_3)$ ;
\item{}
3)\quad
$S_1({\bf n}_3)\rightarrow S_2({\bf n}_4)$ .
\vskip1pc
\item{$iii$)} {\sl Quantum statistics}.
\item{}
The probability of measuring the outcome 1 for
the 1-0 observable represented by the projection operator
$\hat P$ when the state vector is $\psi$ (with $\Vert\psi\Vert=1$),
is given by
$$
p_\psi(\hat P)=\langle \psi\mid\hat  P\psi\rangle.
$$
\par\noindent
Let us suppose that a measurement of $S_1({\bf n}_1)$ yields outcome 1.
Then correlations $(ii)$, together with principle $(i)$, imply that
$S_2({\bf n}_2)$, and hence $S_1({\bf n}_3)$ and $S_2({\bf n}_4)$, are
elements of reality equal to 1. Therefore
conditions $(i)$ and $(ii)$, without using $(iii)$, imply
$$
S_1({\bf n}_1)\;\rightarrow
\;S_2({\bf n}_4).\eqno(1)
$$
From the quantum theoretical point of view, condition $(ii)$ forces the system
in a precise quantum state $\psi$ (see for instance [4]).
Hence we can use such $\psi$ to compute, by $(iii)$, the quantum
probability of obtaining outcomes 1 and 0 from a simultaneous
measurement of
$S_1({\bf n}_1)$ and
$S_2({\bf n}_4)$, respectively.
Such probability
turns out to be different from 0 [1][2].
In other words,
the quantum theoretical prediction contradicts (1). So we have the
following logical situation.
\vskip1pc\noindent
H{\ninerm ARDY's}
T{\ninerm HEOREM} -- $\displaystyle\cases{
(i)\hbox{ and }(ii)\quad\Rightarrow& $(1)$
\hskip23.3mm\hfill{(2i)}\cr
&\cr
(ii)\hbox{ and }(iii)\quad\Rightarrow& not
$(1)$\hskip23.3mm \hfill{(2ii)}\cr}
$
\vskip1pc\noindent
A first important consequence of Hardy's theorem is that
\vskip1pc
\item{I --}
{\sl $(ii)$ and $(iii)$ are not consistent with $(i)$, i.e.
quantum theory is not ``local and realistic''
(in the sense that it does not satisfy $(i)$)}.
\vskip1pc\noindent
No experiment is needed to get such a conclusion, but it is drawn
on a purely theoretical basis.
Another proof of this result has been recently given by Stapp [8];
in such a proof Stapp reaches a contradiction by requiring the
quantum correlations $(ii)$ plus the quantum prediction {\sl not (1)}
(which is not a correlation), but only a {\sl locality}
condition, while the {\sl reality} condition is derived by rigorously
expliciting the {\sl counterfactual} reasoning implicit in $(i)$.
\vskip1pc
On the contrary,
a second, stronger conclusion requires
an experiment performed under
experimental conditions which ensure that
\vskip1pc
\item{(ec)}
{\sl correlations $(ii)$
hold according to quantum theory.}
\vskip1pc
\noindent
It is fair enough that in such experiment a simultaneous
measurement of $S_1({\bf n}_1)$ and $S_2({\bf n}_4)$ respectively
yields outcomes 1 and 0 -- just one time --
(i.e., not (1)), to conclude by (2i) that
\item{II --}
{\sl every empirically valid
theory in which correlations $(ii)$ hold if they hold
according to quantum theory, does not
satisfy the principle of locality and reality $(i)$};
therefore
non-locality must be extended to any
``realistic'' theory which shares correlations $(ii)$ with quantum theory.
\par\noindent
We stress that in the latter argument the experimental test with result
``not (1)'' plays a necessary role. Of course, the occurrence of the
experimental result ``(1)'' would falsify the theory, i.e. it
would be {\sl not empirically valid}.
Moreover, II implies I.
\vskip1pc
The first aim of the present work is to show
that Hardy's type argument, being successful with
respect to conclusion I, fails in reaching II, because the
{\sl required experimental conditions (ec) are not realizable}.
To explicitly see this, we consider
the quantum theoretical description of
the two space-like separated spins,
in the Hilbert space ${\bf C}_1^2\otimes {\bf C}_2^2$,
where ${\bf C}_k^2$ is the Hilbert space for describing the spin
of particle $k$.
Let
$(u_k,v_k)$ be an orthonormal basis of ${\bf C}^2_k$.
An orthonormal basis for the entire space
is $(u_1\otimes u_2, u_1\otimes v_2, v_1\otimes u_2, v_1\otimes v_2)$.
Since any projection operator of ${\bf C}_k^2$ may be written in the form
$E^{\theta,\phi}_k=\left[\matrix{
\cos^2 \theta/2&e^{-i\phi/2}\cos\theta/2\sin\theta/2\cr
e^{i\phi/2}\cos\theta/2\sin\theta/2&\sin^2\theta/2\cr}\right]$
by a suitable choice of the angles $\theta$ and $\phi$, then
the projection operator
${\hat S}_1 ({\bf n})=E_1^{\theta,\phi}\otimes{\bf 1}$
(resp. ${\hat S}_2({\bf n})={\bf 1}\otimes E_2^{\theta,\phi}$)
of ${\bf C}^2_1\otimes{\bf C}^2_2$ represents the
observable $S_1({\bf n})$ (resp. $S_2({\bf n})$), where
${\bf n}=(\sin\theta\cos\phi,\sin\theta\sin\phi,\cos\theta)$.
\vskip1pc\noindent
If $S_1$ and $S_2$ are two measurable together 1-0 observables,
according to quantum theory the correlation
$S_1\to S_2$ holds if and only if the quantum probability
of measuring
1 for $S_1$ and $0$ for $S_2$ is 0, i.e. if and only if
$$
\langle {\hat S}_1({\bf 1}-{\hat S}_2)\psi\mid \psi\rangle=0\quad\hbox{iff}\quad
{\hat S}_1\psi={\hat S}_1{\hat S}_2\psi. \eqno(3)
$$
To emphasize the physical meaning of $(3)$
we rewrite it in the following form,
which explicitly exhibits the state dependence of the correlation.
$$
{\hat S}_1\ipsi {\hat S}_2
$$
Let ${\hat S}_1({\bf n})=F\otimes{\bf 1}$ and ${\hat S}_2({\bf m})={\bf 1}\otimes A$
be two space-like separated 1-0 quantum observables.
If we
choose the basis of ${\bf C}^2_1$ as the basis of the eigenvectors
of $F$, i.e. if $Fu_1=u_1$ and $Fv_1=0$, then $F$ is represented
by the matrix $\left[\matrix{1&0\cr 0&0\cr}\right]$ and
${\hat S}_1({\bf n})=F\otimes{\bf 1}\equiv\left[\matrix{\bf 1&\bf 0\cr\bf 0&\bf 0\cr}
\right]$.
If $\left[\matrix{a\cr b\cr c\cr d}\right]$ represents the vector
state $\psi$, the condition ${\hat S}_1({\bf n})\ipsi{\bf 1}\otimes A$ holds if and only
if $A \left[\matrix{a\cr b}\right]=\left[\matrix{a\cr b}\right]$;
therefore, when
$\left[\matrix{a\cr b}\right]\neq \left[\matrix{0\cr 0}\right]$
(iff ${\hat S}_1({\bf n})\psi\neq 0$), there is a unique projection operator $A$
satisfying such condition, namely
$A={1\over \vert a\vert^2+\vert b
\vert^2}\left\vert\left[\matrix{a \cr b}\right]\bigr\rangle\bigl\langle\left[\matrix{a
\cr b}\right]\right\vert$.
Thus the following statement holds.
\vskip1pc
\noindent
P{\ninerm ROPOSITION} 1. {\sl
 Let $\psi$ be any state vector of the entire system, and ${\hat S}_1({\bf n})=F\otimes{\bf 1}$,
where $F$ is a projection operator of ${\bf C}^2_1$.
Provided that ${\hat S}_1({\bf n})\psi\neq 0$ there is a unique projecion operator
$A={1\over \vert a\vert^2+\vert b
\vert^2}\left\vert\left[\matrix{a \cr b}\right]\bigr\rangle\bigl\langle\left[\matrix{a
\cr b}\right]\right\vert$ of
${\bf C}^2$ such that
${\hat S}_1({\bf n})\ipsi{\bf 1}\otimes A$.   }
\vskip1pc\noindent
Proposition 1 implies that,
once choosen the first direction ${\bf n}_1$ in such
a way that ${\hat S}_1({\bf n}_1)\psi\neq 0$, there is a unique triple of
directions ${\bf n}_2$, ${\bf n}_3$ and ${\bf n}_4$
such that conditions $(ii)$ hold.
An absolute absence of errors in the relative orientations
${\bf n}_1$ and ${\bf n}_4$ in a real experiment is impossible.
Since the existence of an experimental error on ${\bf n}_1$ and
${\bf n}_4$,
whatever be its
entity, provokes the breadown of quantum correlations $(ii)$,
the possibility of an experimental test
of (1) under conditon (ec) is completely hopeless.
For these reasons we cannot reach conclusion (II)
by using Hardy's theorem.
\vskip1pc\noindent
R{\ninerm EMARK. }
Conclusion I, with the strict correlations of the singlet state
instead of (ii), is provided also by Bell's theorem without the need
of experiments.
The necessity of the experiment rises only to get conclusion II.
The experiment required by Bell's argument consists in measuring three
alternative pairs of observables, whose {\sl reality}
is ensured {\sl for all directions}, to check whether
satisfies Bell's inequality [9]. Moreover, according to quantum theory,
the statistical magnitudes
involved in Bell's inequalities are {\sl continuous} functions of
the orientations of the measuring apparatuses. Therefore the
experimental violation of the inequalities is expected
to be guaranteed by
limiting the experimental errors on such orientations within suitable
bounds.
By using Hardy's argument, the much simpler experiment consists in meausuring
only the {\sl two} observables $S_1({\bf n}_1)$ and $S_2({\bf n}_4)$
until the occurrence of the pair of outcomes (1,0) realizes
the violation of (1); unfortunately, such experiment is unrealizable. So,
although Hardy's theorem ``is the best
version of Bell's theorem'' because of its ``highest attainable degree
of simplicity and physical insight'' [3], it suffers a lack of epistemological
efficacy with repect the older Bell's argument.
\vskip1pc
On the contrary, the non-locality theorem without inequalities presented by
GHSZ [7] reaches a conclusion like II without the need of
experiments, but it requires at least three
particles. Here we briefly sketch the argument in the case of four
spin-$1\over 2$ space-like separated particles.
The ingredients of GHSZ's theorem are
\vskip1pc
\item{$i$)} {\sl Principle of locality and reality}
\item{}
The same as for Hardy's theorem.
\vskip1pc\noindent
\item{$ii'$)}{\sl Correlations}
\item{}
A finite set of correlations, each correlation involving
four pairwise space-like separated spin observables. One of the observables
involved in these correlations is $S_1({\bf n}_0)$, i.e. the 1-0
observable describing the spin of the first particle in direction
${\bf n}_0=(1,0,0)$.
\vskip1pc\noindent
Notice that quantum statistics $(iii)$ is not present in these premises.
GHSZ proved that if conditons $(i)$ and $(ii')$ hold, then $S_1({\bf n}_0)$
turns out to be an element of reality.
\par\noindent
The central role in the
argument is played by the following statement.
\vskip1pc\noindent
GHSZ's T{\ninerm HEOREM.} --
$\displaystyle{\cases{
(i)\hbox{ and }(ii')&\hskip-2mm$\Rightarrow\hskip1mm
S_1({\bf n}_0)\to 1-S_1({\bf n}_0)$\hskip8.4mm\hfill{
(4i)}\cr
&\cr
(i)\hbox{ and }(ii')&\hskip-2mm$\Rightarrow\hskip1mm
1-S_1({\bf n}_0)\to S_1({\bf n}_0)$\hskip8.4mm\hfill{
(4ii)}\cr}   }
$
\vskip1pc\noindent
It must be said that (4i), by itself, does not necessarily yield inconsistency:
it is a correlation stating that outcome 1 for $S_1({\bf n}_0)$ cannot
occur. But (4ii) says that outcome 0 is impossible too.
As a consequence,
it turns out impossible to consistently
assign a value to $S_1({\bf n}_0)$, while it is an element of reality.
Therefore, every theory which predicts correlations $(ii')$ is inconsistent
with the principle $(i)$.
Since there is a quantum state $\psi$ for which correlations $(ii')$ do
hold, GHSZ's theorem implies conclusion II (with $(ii')$ replacing $(ii)$),
and no experiment is needed
to get this result.
Thus GHSZ prove the non-locality of any {\sl realistic} theory
which shares with quantum theory the set $(ii')$ of correlations,
without inequalities and without experiments.
However,
their proof holds for systems consisting of at least three particles.
\par
GHSZ's argument suggests
the idea of proving conclusion II for
a system of {\sl two} spin-$1\over 2$ particles,
by following the same logical lines
which avoid the necessity of (unrealizable) experiments.
To realize such a program it is necessary, {\sl at least},
to fulfil the following tasks (a) and (b).
\vskip1pc
\item{a)}{\sl
To find a set ${\cal R}$ of correlations, each correlation involving
two space-like separated
1-0 observables such that from ${\cal R}$ and $(i)$ derives the contradiction
$$
S_1({\bf n}_1)\to 1-S_1({\bf n}_1)\eqno
(5)
$$
\item{}
where $S_1({\bf n}_1)$ is one of the observables involved
in correlations ${\cal R}$.}
\vskip1pc\noindent
To establish the second task, we notice that
according to quantum theory
$$
{\hat S}_1({\bf n}_1)\ipsi {\bf 1}-{\hat S}_1({\bf n}_1)\quad\hbox{iff}\quad
{\hat S}_1({\bf n}_1)\psi=0
$$
and quantum theory by itself, i.e. without further assumptions as $(i)$,
is a {\sl consistent theory}. Then, in order that (5) be a contradiction,
it must be not predicted by quantum theory, i.e. we have to require
${\hat S}_1({\bf n}_1)\psi\neq 0$.
More generally, if
${\hat S}_1({\bf n}_1)\psi= 0$ or
$({\bf 1}-{\hat S}_1({\bf n}_1))\psi= 0$, then $\psi$ is not an entangled
state vector, i.e. it has the form $\varphi_1\otimes\varphi_2$.
Hence no quantum correlation holds between two space-like separated
observables. In this case
the locality and reality principle plays no role and therefore
no contradiction can take place because of it.
Thus, another indispensable task of a GHSZ-type program is
\vskip1pc\noindent
\item{b)}{\sl
To find a quantum state $\psi$ such that
\item{} b.i)
\quad correlations ${\cal R}$ hold according to
quantum theory, and
\item{} b.ii)
\quad ${\hat S}_1({\bf n}_1)\psi\neq 0\neq ({\bf 1}-{\hat S}_1({\bf n}_1))\psi$.}
\vskip1pc\noindent
In the remaining part of our work we show that these two
tasks are unrealizable simultaneously.
\vskip1pc
First we introduce the concept of ``chain'' of correlations, elsewhere
called {\sl ladder} [5].
By {\sl chain} we mean any finite, ordered sequence
${\cal C}=\{S_1({\bf m}_1), S_2({\bf m}_2),S_1({\bf m}_3),
...,S_1({\bf m}_{2k-1}),S_2({\bf m}_{2k})...\}$ of ``local'' 1-0 observables
such that the following chain of correlations holds.
$$
S_1({\bf m}_1)\to S_2({\bf m}_2)\to\cdots
\to S_2({\bf m}_{2k})\to
S_1({\bf m}_{2k+1})\to\cdots
\eqno(6)
$$
\vskip1pc\noindent
L{\ninerm EMMA} 1. -- Let
${\cal C}=\{S_1({\bf m}_1), S_2({\bf m}_2),S_1({\bf m}_3),...\}$ be a chain.
If the state vector $\psi$ is such that (6) hold according to
quantum theory, i.e. if
$$
{\hat S}_i({\bf m}_k)\ipsi {\hat S}_{3-i}({\bf m}_{k+1}),\quad\forall
{\hat S}_i({\bf m}_{k}),
{\hat S}_{3-i}({\bf m}_{k+1})\in{\cal C},
\eqno(7)
$$
then
$$
{\hat S}_1({\bf m}_1)\psi\neq 0\quad\Rightarrow
\quad {\hat S}_i({\bf m}_{k})\psi\neq 0\quad
\forall{\hat S}_i({\bf m}_{k})\in{\cal C}.
\eqno(8)
$$
\vskip1pc\noindent
L{\ninerm EMMA} 2. --
$
{\hat S}_1({\bf n})\ipsi {\hat S}_2({\bf m})\quad\hbox{iff}\quad
{\bf 1}-{\hat S}_2({\bf m})\ipsi{\bf 1}- {\hat S}_1({\bf n}).
$
\vskip1pc\noindent
We do start our argument by considering {\sl any}
finite set ${\cal B}=\{S_r({\bf n}_s)\}$ of
``local'' 1-0 observables, with $S_1({\bf n}_1)\in\cal B$,
endowed with a finite set
${\cal R}=\{[S_1({\bf n}_\lambda)\to
S_2({\bf n}_\rho)]\}$ of correlations. Then we show that there is
no state vector
$\psi$ which satisfies (b.ii) such that
$$\cases{
[S_1({\bf n}_\lambda)\to S_2({\bf n}_\rho)]\in{\cal R}\quad\Rightarrow&
${\hat S}_1({\bf n}_\lambda)\ipsi {\hat S}_2({\bf n}_\rho)$\hskip20.8mm(b.i)\cr
&\cr
{\cal R}\hbox{ and }(i)\quad\Rightarrow& \hskip-25mm$S_1({\bf n}_1)\to 1-S_1({\bf
n}_1)$\hskip42.8mm(a)\cr}
$$
In so doing we do not lose generality. Indeed,
{\sl any correlation} between two measurable together
1-0 observables $S_1$ and $S_2$ consists in nothing else but
the fact that some of the four pairs of outcomes
$(0,0),(0,1),(1,0),(1,1)$ are impossible.
For instance, suppose that (1,0) and (0,1) cannot occur. This
particular correlation is expressed by the two formulas
$S_1\to S_2$ and $S_2\to S_1$.
We assume that the following {\sl obvious} rule must hold in ${\cal R}$.
$$
[S_1({\bf n})\to
S_2({\bf m})]\in{\cal R}\quad\Leftrightarrow\quad
[1 -S_2({\bf m})\to 1-
S_1({\bf n})]\in{\cal R}.\leqno(R1)
$$
\vskip1pc\noindent
According to (b)
we have to assume
${\hat S}_1({\bf n}_1)\psi\neq 0$
and
$({\bf 1}-{\hat S}_1({\bf n}_1))\psi\neq 0$.
The two correlations
${\hat S}_i({\bf n}_\lambda)\ipsi {\hat S}_{3-i}({\bf n}_\alpha)$
and ${\hat S}_i({\bf n}_\lambda)\ipsi {\hat S}_{3-i}({\bf n}_\beta)$
imply
${\hat S}_{3-i}({\bf n}_\alpha)=
{\hat S}_{3-i}({\bf n}_\beta)$ (by prop.1 and lemma 1). Therefore in $\cal B$
there is a unique maximal chain ${\cal C}_1$ containing
${S}_1({\bf n}_1)$
and a unique maximal chain ${\cal C}_0$ containing
$(1-{S}_1({\bf n}_1))$.
This means that {\sl there is no correlation in ${\cal R}$ between any observable
in ${\cal C}_1$ or in ${\cal C}_0$ and any other observable in
${\cal B}\setminus({\cal C}_1\cup{\cal C}_0)$.}
Therefore, the correlation
$S_1({\bf n}_1)\to 1-S_1({\bf n}_1)$
can be derived by using $(i)$ only within the correlations
in the chains ${\cal C}_1$ or ${\cal C}_0$.
\vskip1pc\noindent
Now, for {\sl any} pair $(S_i({\bf n}_k),S_j({\bf n}_h))$ of measurable
together observables in the same chain $\cal C$
(i.e. such that $[{\hat S}_i({\bf n}_k),{\hat S}_j({\bf n}_h)] = {\bf 0}$),
either
$S_i({\bf n}_k)\to S_j({\bf n}_h)$ or
$S_j({\bf n}_h)\to S_i({\bf n}_k)$ can be always derived
by using
principle $(i)$ like we have done to get (1).
Therefore, within a chain $\cal C$
the only rule provided by $(i)$ for deriving new correlations other than
those provided by ${\cal R}$ is the following
$$
S_i({\bf n}_k)\to S_j({\bf n}_h)\quad\hbox{iff}\quad k\leq h.\leqno(R2)
$$
As a consequence,
the correlation
$S_1({\bf n}_1)\to 1-S_1({\bf n}_1)$ can be derived from  ${\cal R}$ and $(i)$
only if in ${\cal C}_1$ the observable
$S_1({\bf n}_1)$ precedes $1-S_1({\bf n}_1)$.
This may happen only if there exists $S_2({\bf n}_{2k})$ in the chain ${\cal C}_1$
such that $S_2({\bf n}_{2k})\to
1-S_1({\bf n}_1)$. But the following proposition 2
states that such $S_2({\bf n}_{2k})$
cannot exist whenever $S_1({\bf n}_1)\psi\neq 0$.
\vskip1pc\noindent
P{\ninerm ROPOSITION} 2. -- Let
${\cal C}=\{S_1({\bf n_1}), S_2({\bf n}_2),S_1({\bf n}_3),...\}$ be a chain
such that $S_2({\bf n}_{2k})\to 1-S_1({\bf n}_1)$ for some $k$.
If there
is $\psi\in{\bf C}^2\otimes{\bf C}^2$ such that correlations (6)
hold according to quantum theory, then
${\hat S}_1({\bf n}_1)\psi=0$.
\vskip1pc\noindent
P{\ninerm ROOF.} Let $\psi$ be a vector state such that (6) hold according to quantum
theory. Let $k$ be such that
$$
{\hat S}_2({\bf n}_{2k})\ipsi {\bf 1}-{\hat S}_1({\bf n}_1)\eqno(9)
$$
We prove that ${\hat S}_1({\bf n}_1)\psi=0$. Indeed, if
${\hat S}_1({\bf n}_1)\psi\neq 0$,
since ${\hat S}_2({\bf n}_{2k})\ipsi {\hat S}_1({\bf n}_{2k+1})$ and (9) hold,
proposition 1 and lemma 1 imply
${\bf 1}-{\hat S}_1({\bf n}_1)={\hat S}_1({\bf n}_{2k+1})$, i.e.
$$
{\hat S}_1({\bf n}_1)=
{\bf 1}-{\hat S}_1({\bf n}_{2k+1}).\eqno(10)
$$
If ${\hat S}_1({\bf n}_1)\psi=\psi$, then from (10) we have
${\hat S}_1({\bf n}_{2k+1})\psi=0$, contrary to lemma 1. Then
$0\neq {\hat S}_1({\bf n}_1)\psi\neq \psi$ and
$0\neq {\bf 1}-{\hat S}_1({\bf n}_1)\psi\neq \psi$ hold. More generally, using
lemmas 1 and 2, it can be proved that
$$
0\neq {\hat S}_r({\bf n}_s)\psi\neq \psi\quad\hbox{and}\quad
0\neq ({\bf 1}-{\hat S}_r({\bf n}_s))\psi\neq \psi
\eqno(11)
$$
hold for all $S_r({\bf n}_s)\in\cal C$.
Since
$$
\cases{
{\hat S}_1({\bf n}_1)\ipsi &\hskip-11mm${\hat S}_2({\bf n}_2)$\cr
&\cr
{\bf 1}-{\hat S}_1({\bf n}_{2k+1})\ipsi &${\bf 1}-{\hat S}_2({\bf n}_{2k})$,\cr
}$$
(10) and prop.1 imply
$$
{\hat S}_2({\bf n}_2)={\bf 1}-{\hat S}_2({\bf n}_{2k}).
$$
By iterating this argument, making use of prop.1 and (11), we get
$$
{\hat S}_{i(j)}({\bf n}_{1+j})={\bf 1}-
{\hat S}_{i(j)}({\bf n}_{2k+1-j}),
$$
for all $j=1,2,...,2k$, where $i(j)={3-(-1)^j\over 2}\in\{1,2\}$
is the appropriate index.
In particular, for $j=k$ we get the ``impossible'' equation
$$
{\hat S}_{i(k)}({\bf n}_{k+1})={\bf 1}-
{\hat S}_{i(k)}({\bf n}_{k+1}).
\eqno(12)
$$
Prop. 2 completes our argument against the possibility of proving
non-locality of a realistic theory, describing two space-like separated
two-level sub-systems, which shares a set
of correlations with quantum theory,
by using a GHSZ type -- without inequalities and without experiment --
method.
\vskip3pc\noindent
{\bf References.}\par\noindent
[1] L. Hardy, Phys.Rev.Lett. {\bf 71}, 1665 (1993).
\par\noindent
[2] S. Goldstein, Phys.Rev.Lett. {\bf 72}, 1951 (1994).
\par\noindent
[3] N.D. Mermin, Ann. N.Y. Acad. Sci. {\bf 755}, 616 (1995).
\par\noindent
[4] G. Kar, Phys.Rev. A {\bf 56}, 1023 (1996); J.Phys. A. {\bf 30},
217 (1997).
\par\noindent
[5] A. Cabello, Phys.Rev. A {\bf 58}, 1687 (1997).
\par\noindent
[6] J.S. Bell, Physics, {\bf 1}, 165 (1965).
\par\noindent
[7]
D.M. Greenberger, M.A. Horne, A. Shimony and A. Zeilinger,\par
Am.J.Phys., {\bf 58}, 1131 (1990).
\par\noindent
[8] H.P.Stapp, {\sl Nonlocality, counterfactuals,
and consistent histories} \par
LBNL-43201, University of California,
Berkeley 1999.
\par\noindent
[9] A. Aspect, in {\it Conference Proceedings of Italian Physical Society},
\par Vol. {\bf 60}, p.345,
R. Pratesi and L. Ronchi eds.,
(Compositori, \par Bologna 1998).
\bye